\documentclass[pra,twocolumn,showpacs,superscriptaddress]{revtex4-1}

\usepackage{graphicx}
\usepackage{setspace}
\usepackage{amsmath}
\usepackage{mathtools}
\usepackage{tabularx}
\usepackage{mathrsfs}
\usepackage{xcolor}
\usepackage[citecolor=blue,linkcolor=blue,colorlinks=true,linkbordercolor=white]{hyperref}

\newcommand{\be}{\begin{equation}}
\newcommand{\ee}{\end{equation}}
\newcommand{\bea}{\begin{eqnarray}}
\newcommand{\eea}{\end{eqnarray}}

\newcommand{\la}{\langle}
\newcommand{\ra}{\rangle}
\newcommand{\ua}{\uparrow}
\newcommand{\da}{\downarrow}

\newcommand{\bigp}[1]{
	\left( #1 \right)
}

\begin{document}

\title{Entanglement in disordered superfluids: the impact of density, interaction and harmonic confinement on the Superconductor-Insulator transition}

\author{G. A. Canella}
\affiliation{Institute of Chemistry, S\~{a}o Paulo State University, 14800-090, Araraquara, S\~{a}o Paulo, Brazil}
\author{V. V. Fran\c{c}a}

\affiliation{Institute of Chemistry, S\~{a}o Paulo State University, 14800-090, Araraquara, S\~{a}o Paulo, Brazil}

\begin{abstract}

We investigate the influence of density, interaction and harmonic confinement on the superfluid to insulator transition (SIT) in disordered fermionic superfluids described by the one-dimensional Hubbard model. We quantify the ground-state single-site entanglement via density-functional theory calculations of the linear entropy. We analyze the critical concentration $C_C$ at which the fully-localized state $-$ a special type of localization, with null entanglement $-$ emerges. We find that $C_C$ is independent on the interaction, but demands a minimum disorder strength to occur.  We then derive analytical relations for $C_C$ as a function of the average particle density for attractive and repulsive disorder. Our results reveal that weak harmonic confinement does not impact the properties of the fully-localized state, which occurs at the same $C_C$, but stronger confinements may lead the system from the fully-localized state to the ordinary localization.

\end{abstract}

\pacs{}

\maketitle

\section{Introduction}

Localization, since it was first modeled by Anderson \cite{anderson,anderson2,anderson3}, has been investigated in several disordered 
systems, from theoretical works, exploring its 
properties and conditions under which it emerges, to experimental observations, specially in ultracold atomic gases. 

For localization occurring in disordered superfluids one of the goals is to understand the so-called superfluid to insulator transition 
(SIT): in which superfluids are transformed into insulators under strong or moderate disorder. A broad comprehension of the SIT implies also 
a better understanding of several complex superconductors  
\cite{natureRef3,benjaminRef1-2,benjaminRef1-1,benjaminRef1-4,LyeBoseEins,sundarSciRep}, including for example high-$T_c$ superconductors. 

Entanglement $-$ one of the crucial ingredients for the development of future quantum technologies \cite{acin} $-$ has been recognized 
as a powerful tool for detecting quantum phase transitions and crossovers in several contexts \cite{qpt1,qpt2,qpt3}. There are several well defined entanglement measures, as for example the von Neumann entropy, which has been used for quantifying bipartite 
entanglement of pure states \cite{AmicoMan}. The so-called single-site entanglement \cite{zanardi} $-$ defined as the entanglement between 
a single-site of a discrete model and the remaining sites $-$ has been explored in the homogeneous Hubbard model and associated to quantum 
phase transitions \cite{vvf2006,oster,gu1,larsson1}. 

A local-density approximation (LDA) for the entanglement entropy of any generic inhomogeneous system has been proposed \cite{vvf2008prl} and successfully applied to the Hubbard \cite{vvf2008prl} and the Kondo \cite{kondo} models, within density-functional theory calculations  \cite{dft1, dft2,dft3,dft4}. Block-block entanglement has also been investigated in 
connection to quantum phase transitions \cite{gu2}, from the view point of its universal and non-universal contributions \cite{vvf2008pra} and from the perspective of engineering 
strongly entangled superlattices \cite{tobi}. 

For highly complex systems $-$ with very many degrees of freedom $-$ a practical alternative to the von Neumann entropy has been proposed: 
the linear entropy \cite{buscemi2007,irene2008,shirwan2009}, which indicates the number and spread of terms in the Schmidt decomposition 
of the state. In general the SIT properties should not depend on which quantity is used to track the transition \cite{CarterMacK}, however some 
measures of entanglement may not be sensitive enough to detect any distinct behavior when passing from the superfluid to the insulator phase. 

In particular, entanglement has been used to investigate SIT in several systems, including spinless 
fermions, Bose gases, mixtures of bosons and fermions 
\cite{berkovitsRef13,berkovitsRef13-7,berkovitsRef13-8,vettRef5,wangRef21,islamRef18,florRef19,goldRef29,dengRef32,frerotRef35,albusRef37,royRef14,yeRef15,yeRef15-25,vvf2011pra}, and very recently, purely fermionic superfluids \cite{canella2019prl}. The linear entropy has been proved to contain remarkable signatures of the SIT in purely fermionic superfluids and a special type of localization was found to emerge for sufficiently strong disorder at a certain critical concentration $C_C$ or at a certain particle density $n_C$ \cite{canella2019prl}. However the impact of the density, the interaction and the harmonic confinement $-$ necessary to 
properly describe the trap in ultracold atoms' experiments $-$ on the aforementioned results remain to be investigated.

We here study the effects of density, interaction and harmonic confinement on the SIT of one-dimensional disordered fermions as 
described by the Hubbard model. In particular, we analyze the critical concentration $C_C$ at which the fully-localized state emerges. We analytically obtain the relations between $C_C$ and the particle density for attractive and repulsive disorder. Our results reveal that while $C_C$ is independent on the interaction, a minimum disorder strength is required to the existence of the full localization.  We find that the fully-localized state occurs at the same $C_C$ for weak harmonic confinement, but stronger confinements may drive the system from the full localization to the ordinary localization.

\section{The Model}

Our disordered superfluids are described by the fermionic Hubbard model \cite{hubb}, 

\be{}
	\hat{H} = -t\sum_{\la ij\ra \sigma}\bigp{\hat c^{\dagger}_{i\sigma}\hat c_{j\sigma} } + 
	U\sum_i \hat{n}_{i\ua}\hat{n}_{i\da} + \sum_{i\sigma}V_i\hat{n}_{i\sigma},
	\label{eq:HubbardHamiltonian}
\ee{}
where $\hat{n}_{i\sigma} = \hat{c}^{\dagger}_{i\sigma}\hat{c}_{i\sigma}$ is the density operator at site $i$ with $z$-spin component 
$\sigma = \ua,\da$,  $\hat{c}^{\dagger}_{i\sigma}$ ($\hat{c}_{i\sigma}$) is the creation (annhilation) fermionic operator, $U$ is the on-site attractive interaction, $t$ the inter-site hopping parameter and $V_i$ represents the external potential. Here we consider chains with size $L = 100$ and average density $n = N/L$, where $N= N_{\ua} + N_{\da}$ is the total number of particles, which is spin-balanced ($N_{\ua} = N_{\da}$). 

The external potential $V_i$ describes pointlike disorder, which are localized impurities randomly distributed along the chain within a 
certain concentration $C$, defined as the percentage relation between the number of sites with impurities $L_V$ and the total number of 
sites, $C = 100L_V/L$. We generate $M = 100$ samples for each set of parameters and then any quantity is analyzed as the average 
over these samples. This procedure is essential to ensure that the results are not dependent on specific configurations of impurities. It 
implies however in a huge amount of data which would be impossible to be obtained with exact methods, such as density-matrix 
renormalization group (DMRG) \cite{DMRG} calculations. 

\begin{figure}[!t]
	\centering

	\includegraphics[scale=0.55]{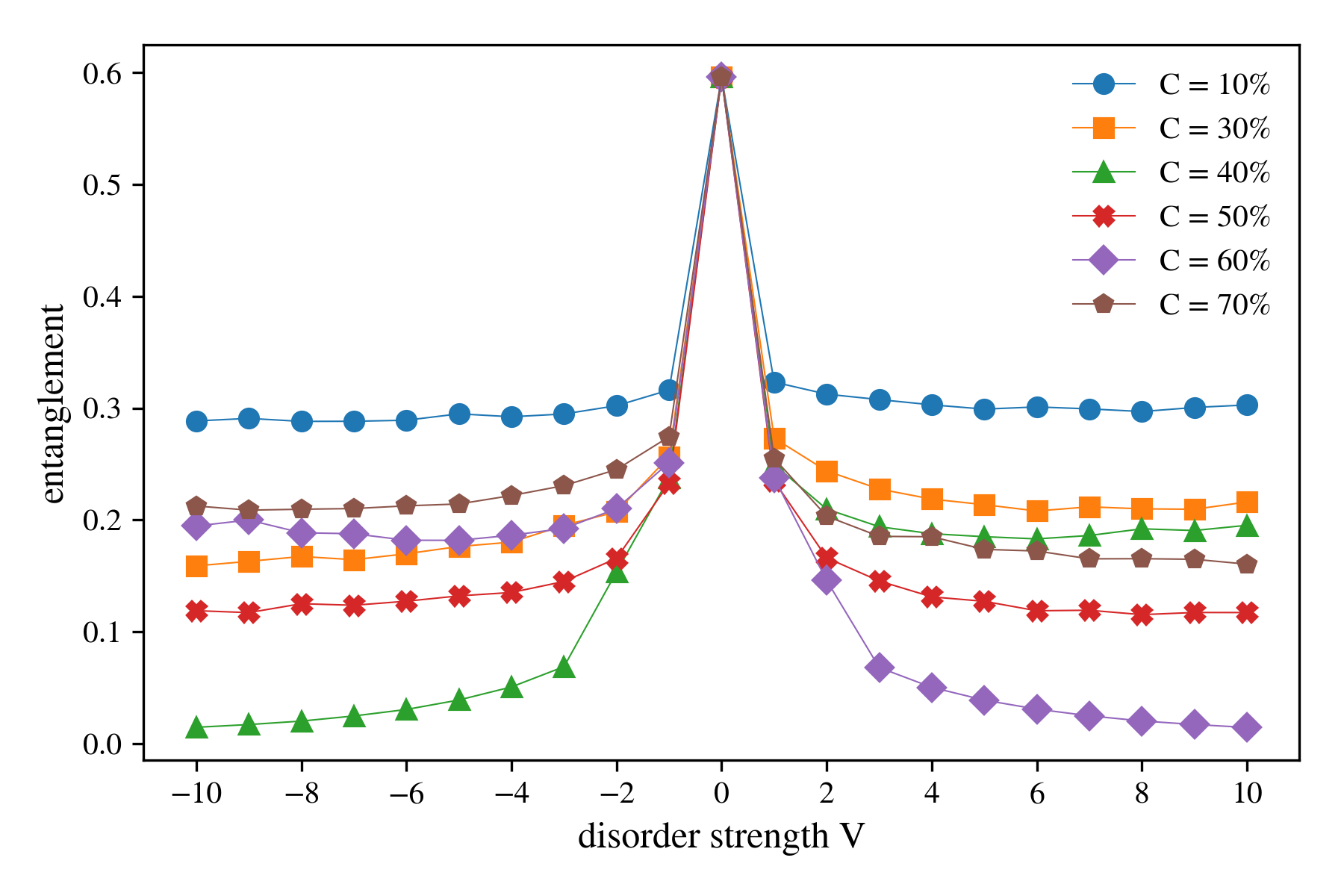}

	\caption{Average single-site ground-state entanglement of disordered superfluid chains, quantified by the linear entropy $\mathcal{L}^{LDA}$, as a function of the disorder strength $V$, for attractive and repulsive disorder, for several concentrations of impurities. Here the on-site interaction is $U=-5t$, the average density is $n=0.8$ and we have adopted open boundary conditions. 
	}

	\label{fig1}
\end{figure}

\begin{figure}[!h]
	\centering
	\includegraphics[scale=0.55, trim = 0mm 12mm 0mm 0mm, clip]{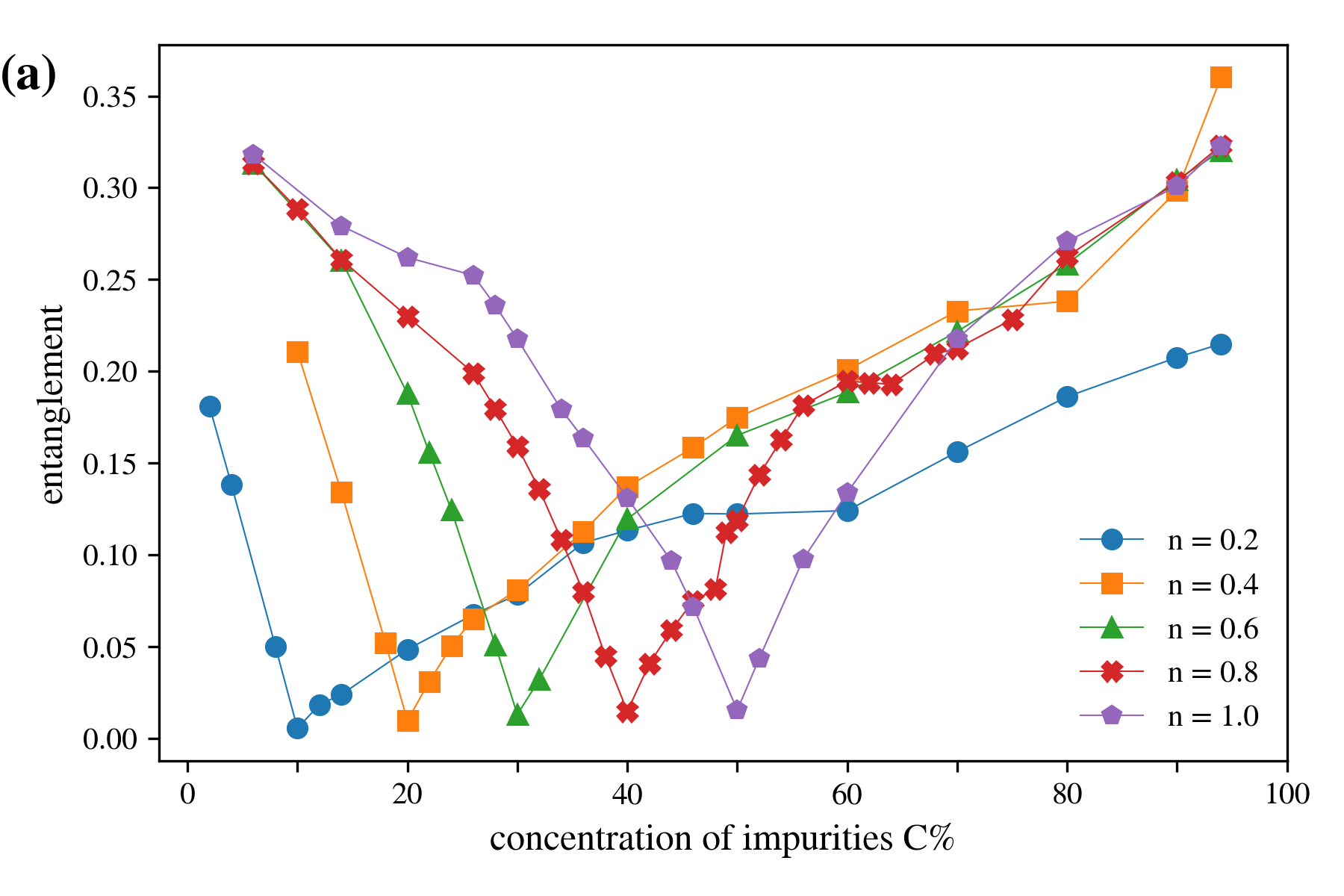}
	\includegraphics[scale=0.55]{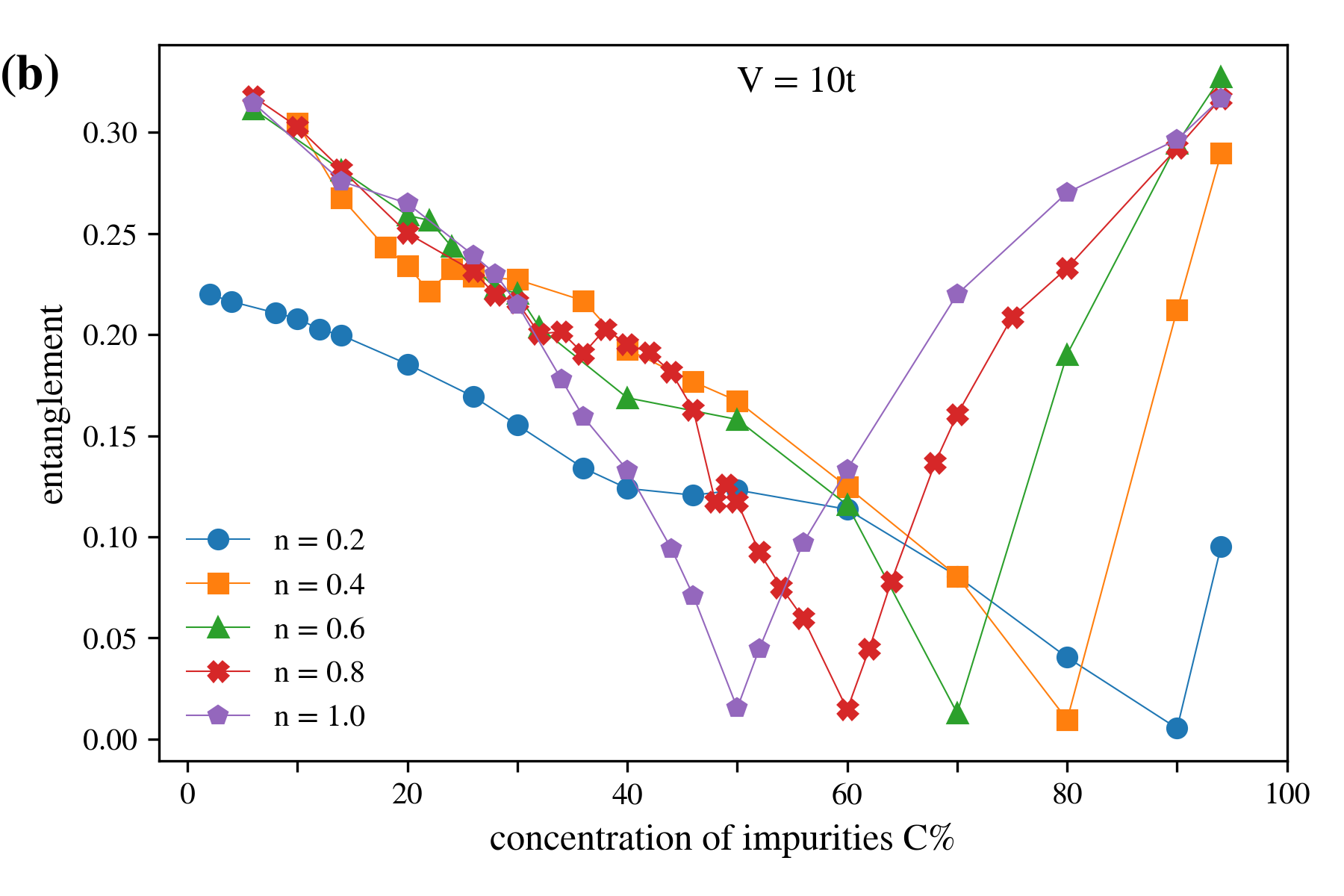}
	\caption{Average single-site ground-state entanglement of disordered superfluid chains, quantified by the linear entropy $\mathcal{L}^{LDA}$, as a function of the impurities' concentration $C$ for (a) attractive disorder ($V=-10t$) and (b) repulsive disorder ($V=10t$), for several densities. Here the on-site interaction is $U=-5t$ and we have adopted open boundary conditions.}

	\label{fig2}
\end{figure}

Our approach consists instead of obtaining an approximated solution via standard density-functional theory (DFT) for the Hubbard model 
\cite{hubb}. We solve the Kohn-Sham cycle for the model, using the fully numerical Bethe-Ansatz solution \cite{liebWu}, which will 
compose the exchange-correlation energy functional (together with the Hartree energy and the single-particle kinetic energy) for 
homogeneous systems. This functional is then used as input within a local-density approximation (LDA) in order to obtain the total energy 
and the density profile of the disordered chains. 

For quantifying the degree of the ground-state entanglement between a single site and the remaining sites, we use a specially designed density functional 
for the linear entropy of the homogeneous Hubbard model \cite{vvf2011pra}

\bea{}
	\mathcal{L}^{hom}(n,U<0) \approx&& \hspace{0.1cm}n - \frac{n^2}{2} + 2\alpha(|U|)\sin\bigp{\frac{\pi n}{2}}	\nonumber\\
				& &- 	  4\alpha^2(|U|)\sin^2\bigp{\frac{\pi n}{2}},
\eea{}
where $\alpha(|U|)$ is given by
\be
\alpha(|U|)=2\int_0^\infty{\frac{J_0(x)J_1(x)e^{|U|x/2}}{(1+e^{|U|x/2})^2}dx},\label{w2n1}
\ee
and $J_\kappa(x)$ are Bessel functions of order $\kappa$. This functional is then used as input in a local-density approximation, following the original LDA protocol which has been proposed for any entanglement 
measure \cite{vvf2008prl}. Thus the linear entropy for each disordered sample is given by
\bea{}
	\mathcal{L}^{inh} &\approx& \mathcal{L}^{LDA}\equiv \frac{1}{L}\sum_i\mathcal{L}^{hom}(n,U<0)|_{n\rightarrow n_i}
\eea{}

\section{The impact of density}

For attractive disorder it has been recently shown \cite{canella2019prl} that the SIT driven by the potential strength $V$ does not require any critical disorder intensity: any small $V$ is enough to drive the transition. We thus start our analysis by investigating the existence or not of a critical intensity for the SIT driven by {\it repulsive} disorder.

\begin{figure}[!t]
	\centering
	\includegraphics[scale=0.42, trim = 0mm 12mm 0mm 0mm, clip]{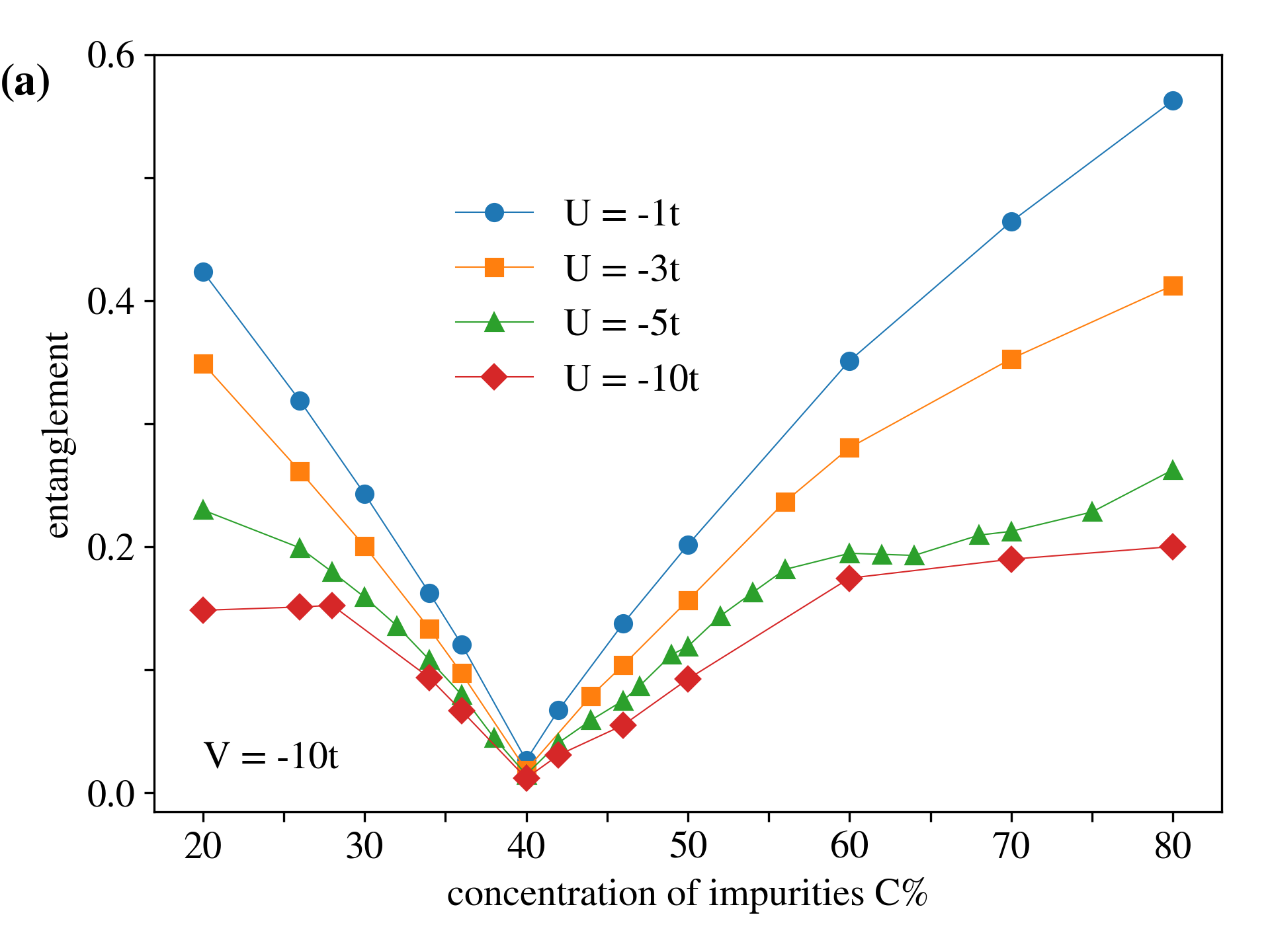}
	\includegraphics[scale=0.42, trim = 0mm 12mm 0mm 0mm, clip]{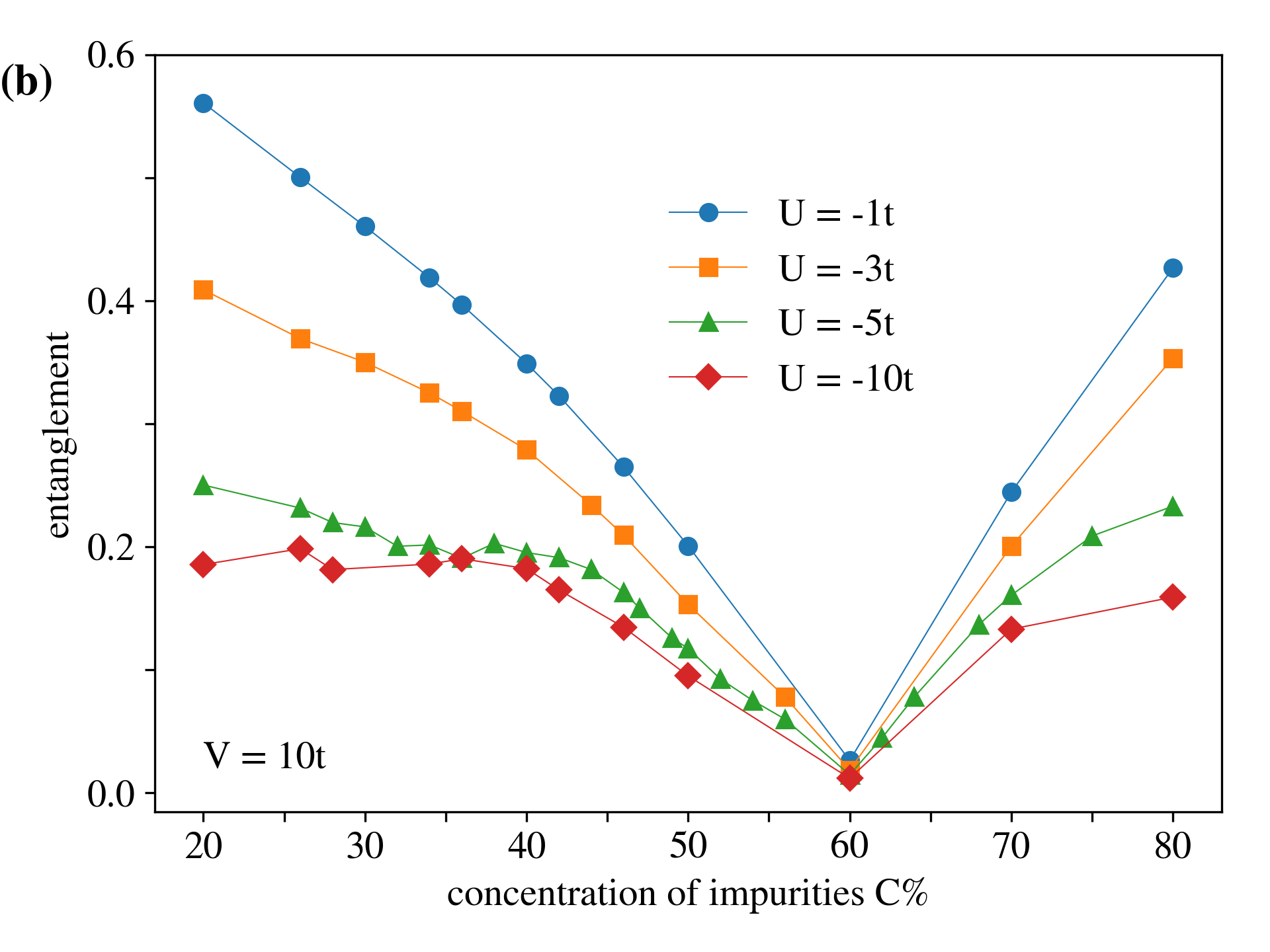}
	\includegraphics[scale=0.42, trim = 0mm 12mm 0mm 0mm, clip]{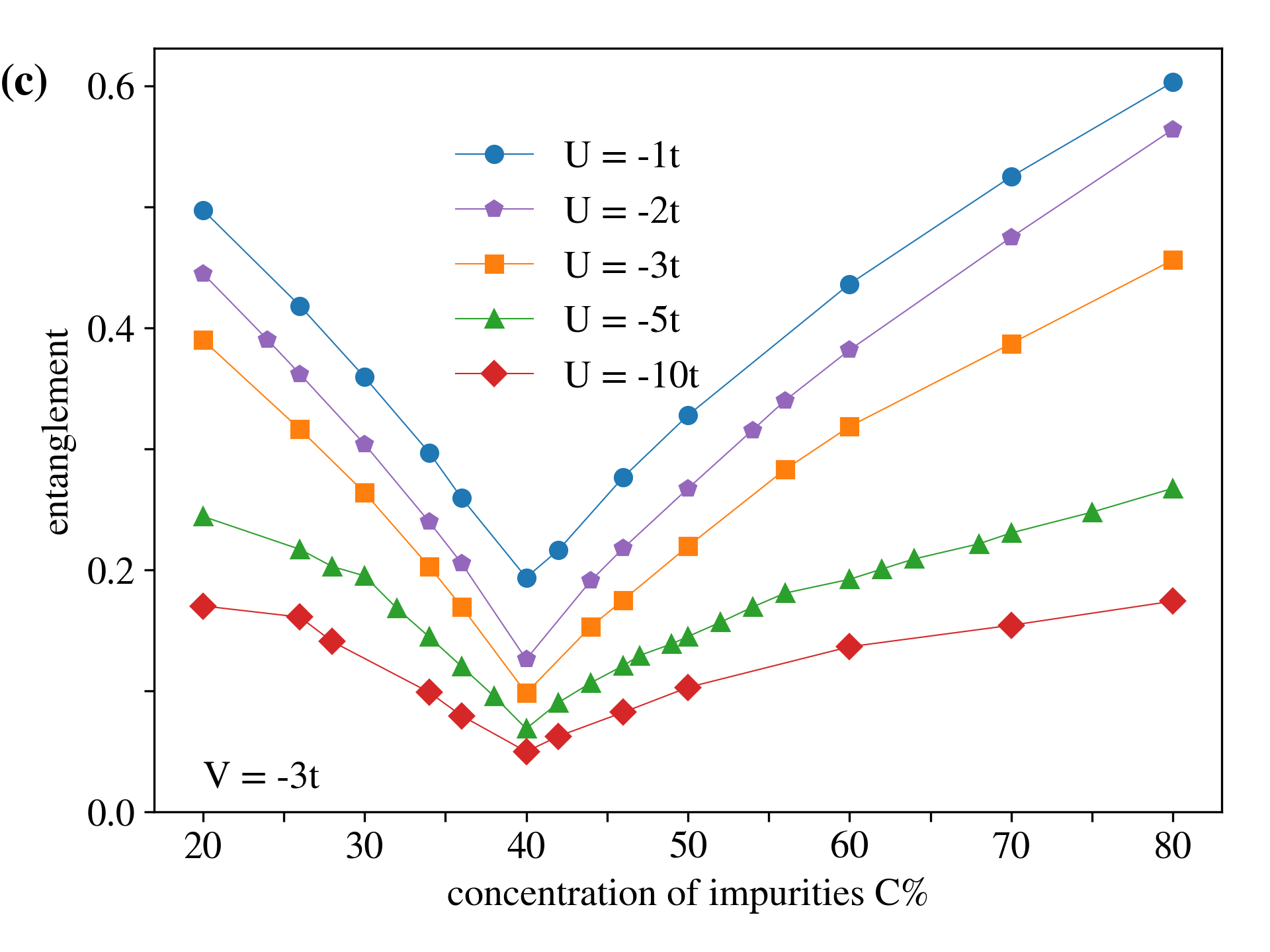}
	\includegraphics[scale=0.42]{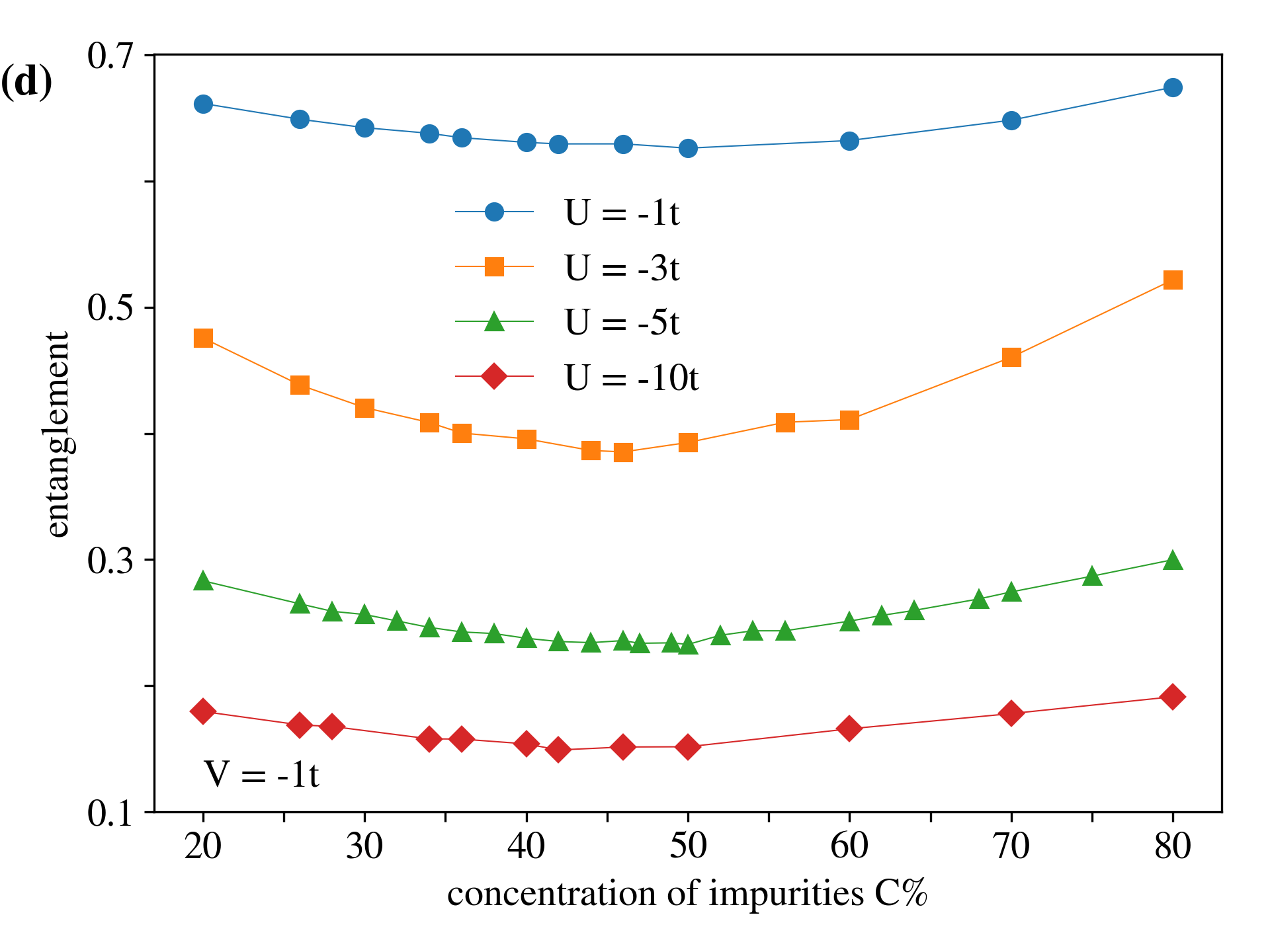}
	\caption{Average single-site ground-state entanglement of disordered superfluid chains, quantified by the linear entropy $\mathcal{L}^{LDA}$, as a function of the disorder concentration $C$ for strong (a,b), moderate (c) and weak (d) disorder strength, for several on-site attractive interactions $U$. In all cases the average density is $n=0.8$ and we have adopted open boundary conditions.}
	\label{fig4}
\end{figure}

Figure \ref{fig1} shows the average single-site entanglement quantified by the linear entropy $\mathcal{L}$ as a function of the disorder strength $V$ for attractive and repulsive disorder and several impurities' concentration. Our results confirm that, similarly to attractive disorder, any small intensity of positive $V$ is enough to decrease considerably the degree of entanglement. Further enhancement of $V$ has almost no impact on the degree of entanglement. This fast entanglement saturation is a clear signature of the SIT without any critical disorder strength. Thus there is no critical disorder intensity for the SIT driven neither 
by attractive nor by repulsive impurities. 

Alike the attractive case, we see in Fig. \ref{fig1} that entanglement is non-monotonic with the concentration, however 
while for $V<0$ the minimum entanglement occurs at a certain critical concentration $C=C_C=40\%$ (for $n=0.8$), for $V>0$ we find the minimum at $C=C_C=60\%$. This reflects the 
reversed role between sites with and without impurities due to the reflection symmetry with respect to $V=0$: $40\%$ of attractive-impurity 
sites are equivalent to $60\%$ of repulsive-impurity sites. This symmetry can be also seen in Fig. \ref{fig1} between attractive $30\%$ and repulsive $70\%$.

Next we explore the dependence of this critical concentration with the average particle density. In Figure \ref{fig2} we present entanglement as a function of concentration for strongly attractive (Fig. \ref{fig2}-a) and strongly 
repulsive (Fig. \ref{fig2}-b) disorder, for several average densities. For this strong disorder regime we find a critical concentration $C_C$ at 
which the entanglement is null, for any average density. Our results show that the $C_C$ occurs when the number of coupled pairs, $N/2$, 
coincides with the number of most favorable sites, i. e. impurity sites $L_V$ for attractive impurities and non-impurity sites $L_0=L-L_V$ 
for repulsive impurities. Thus we find that the critical concentration depends on density as

\begin{eqnarray}
{\text for}\hspace{0.2cm} V<0 &:&
\frac{N}{2}=L_V\nonumber\\
&&\hspace{0cm} \frac{nL}{2}=\frac{CL}{100}\nonumber\\
&&C_C=100\frac{n}{2},\label{eq1}
\end{eqnarray}

\begin{eqnarray}
\hspace{1.5cm}{\text for}\hspace{0.2cm} V>0 &:&
\frac{N}{2}=L_0\nonumber\\
&&\hspace{0cm} \frac{nL}{2}=L-L_V=L-\frac{CL}{100}\nonumber\\
&&C_C=100\left(1-\frac{n}{2}\right).\label{eq2}
\end{eqnarray}

We interprete this as follows: for $C<C_C$ and $V<0$ ($V>0$), as the number of pairs are larger than the number of impurities sites
(non-impurities sites), the pairs spread over the entire chain, despite the fact that the impurities (non-impurities) are the most 
attractive (less repulsive) sites. At $C=C_C$ the strongly coupled dimers fit exactly to the impurity (non-impurity) sites for $V<0$ ($V>0$) 
and thus the system is in a fully-localized state with no entanglement (actually $\mathcal{L}\rightarrow 0$ for $|V|\rightarrow \infty$) \cite{canella2019prl}. For $C>C_C$ the dimers are all in impurity (non-impurity) sites, but as the most favorable sites are in larger number than the dimers population, the system maintains a certain degree of freedom, essentially due to the competition between double-occupation and zero-occupation probabilities. So in this case there is no full localization, but only the ordinary localization, in which entanglement saturates at a finite value $\mathcal{L}>0$.

\begin{figure}
	\centering
	\includegraphics[scale=0.5]{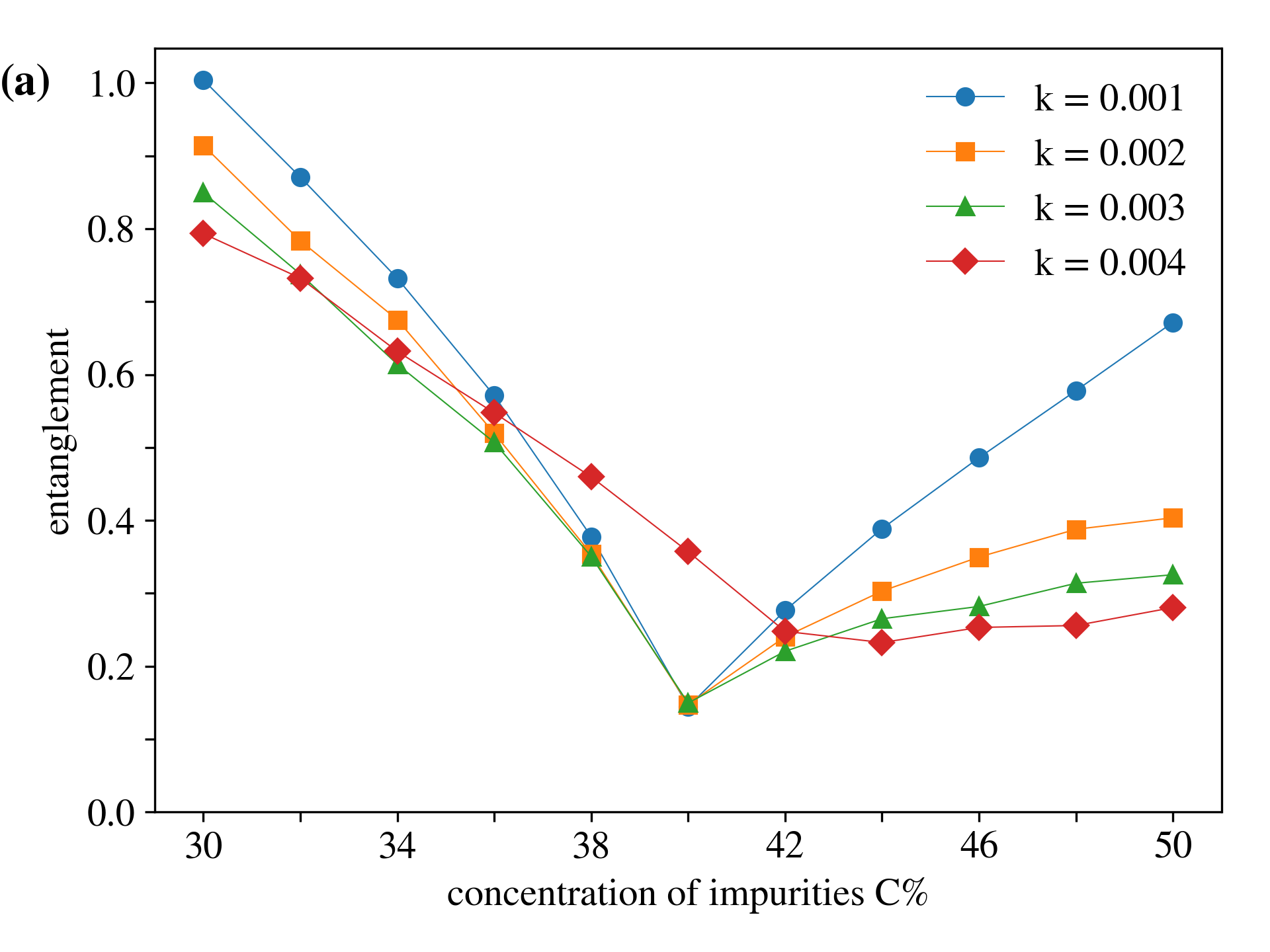}
	\includegraphics[scale=0.5]{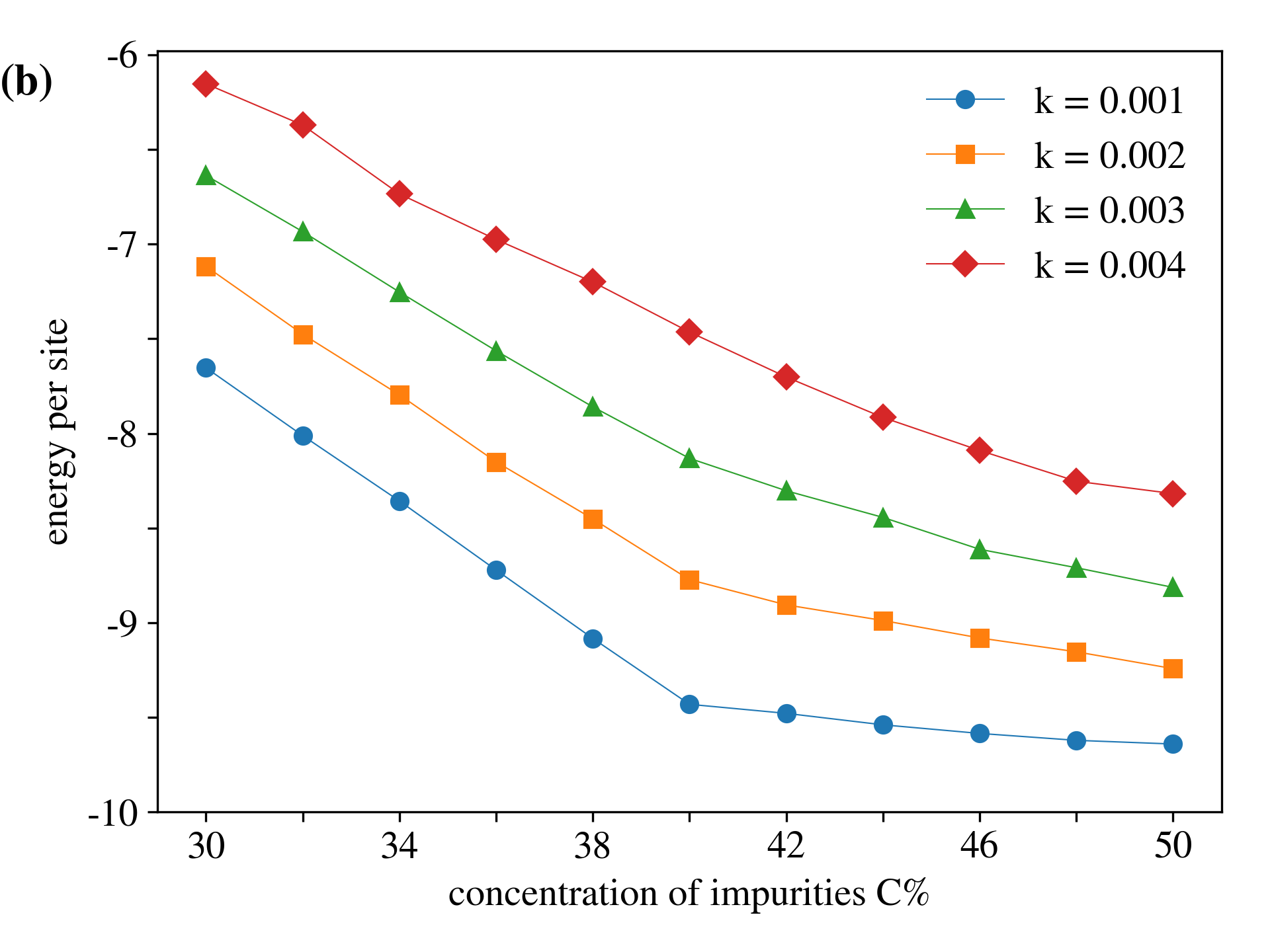}
	\includegraphics[scale=0.5]{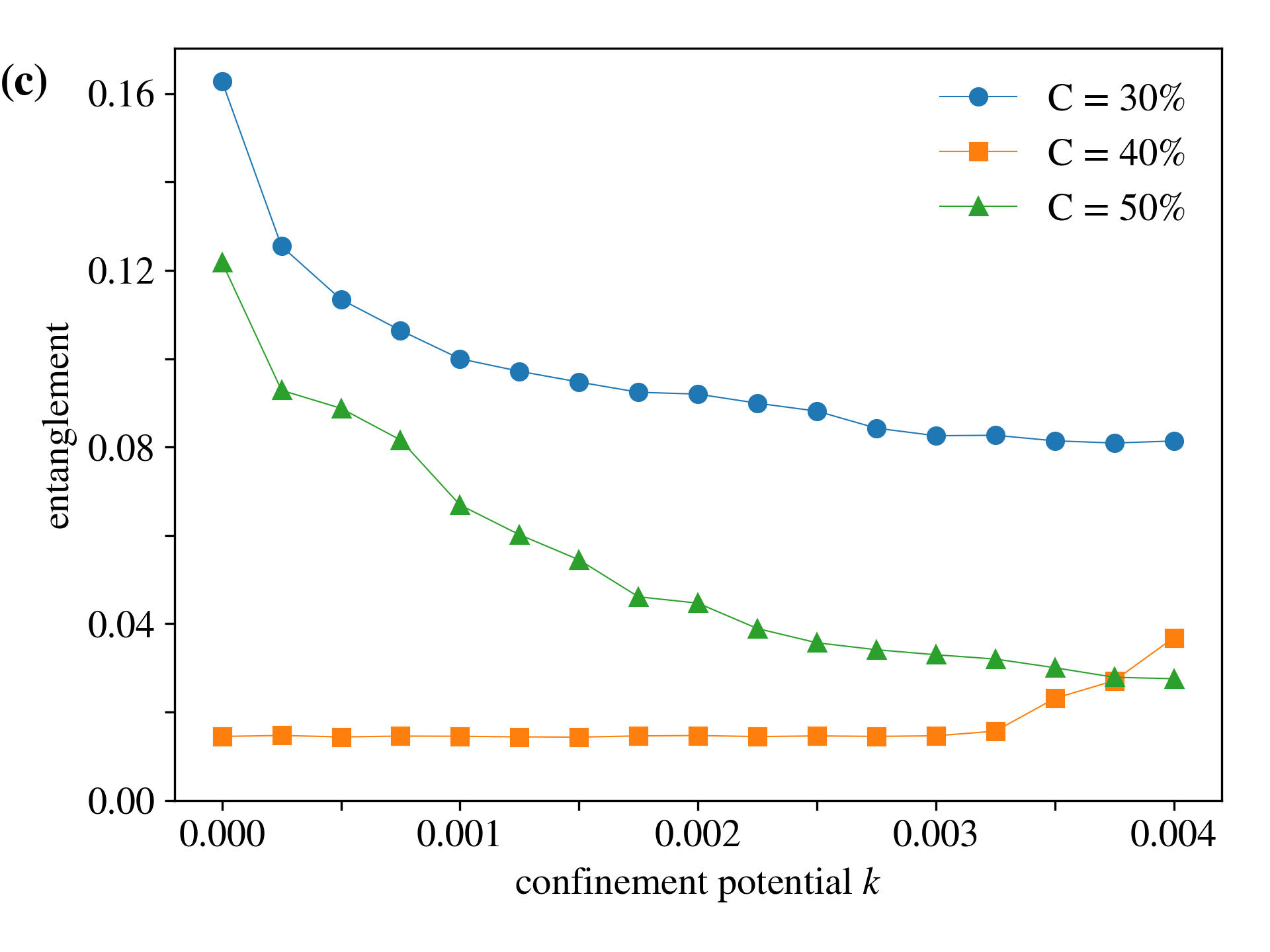}
	\caption{(a) Average single-site ground-state entanglement, quantified by the linear entropy $\mathcal{L}^{LDA}$, and (b) per-site ground-state energy, as a function of the disorder concentration $C$ for harmonically confined superfluid systems, $k(i-i_0)^2$, whose intensity is defined by the curvature $k$ of the harmonic trap. (c) $\mathcal{L}^{LDA}$ as a function of $k$ for $C<C_C$, $C=C_C=40\%$ and $C>C_C$. In all cases the on-site interaction is $U=-5t$, the attractive disorder intensity is $V=-10t$, the particle density is $n=0.8$ and we have adopted open boundary conditions.}
	\label{fig5}
\end{figure}

\section{The impact of interaction}

Another important question to be investigated is how the on-site interaction $U$ impacts the critical concentration for the full localization and the SIT in general. In our superfluids systems, one may imagine that stronger attractive interactions could either reinforce or attenuate the SIT depending on the interplay between $U<0$ and $V$, i.e. if they compete (for $V>0$) or contribute (for $V<0$). Therefore we explore entanglement as a function of the concentration for all the combined regimes: weak, moderate and strong disorder, for weakly, moderate and strongly interacting superfluids. 

Surprisingly though we find 
in Figures \ref{fig4}-a and \ref{fig4}-b, for strong disorder, that $U$ has no impact on $C_C$ neither for attractive nor for repulsive disorder: the critical concentration 
for full localization remains $C_C=100(n/2)\%$ (for $V\rightarrow -\infty$) and $C_C=(1-n/2)100\%$ (for $V\rightarrow \infty$). 

For moderate disorder, Figure \ref{fig4}-c reveals that although entanglement still has a minimum at $C_C$, it is significantly larger than zero for weak on-site interaction $U=-t$. Thus for moderate disorder, there is a minimum on-site interaction, $U_{min}\sim -3t$ in this case, necessary to the system to reach the fully-localized state at $C_C$. 

On the other hand, for weak disorder ($V=-t$), shown in Fig. \ref{fig4}-d, even a very strong interaction, as $U=-10t$, is not enough to lead the system to the fully-localized state. This is due to the fact that for small $V$ the impurity sites are essentially equally favorable to the dimers as the non-impurity ones, so the full localization, characterized by all the dimers at impurity sites, does not occur. All the above results clearly show that the disorder strength dominates the emergence or not of the full localization at the SIT. 

\section{The impact of harmonic confinement}

Finally we analyze the influence of harmonic traps $-$ which are essential in state-of-the-art experiments with ultracold atoms for investigating SIT $-$ on the critical 
concentration $C_C$ for full localization. We then consider in our Hamiltonian (Eq.\ref{eq:HubbardHamiltonian}) two external potentials, the pointlike disorder with a certain concentration $C$ of impurities of intensity $V$ and the parabolic potential $V_i=k(i-i_0)^2$ of curvature $k$, centered at 
$i_0=50.5$ in chains of size $L=100$. 

Figure \ref{fig5}-a shows that for $k\lesssim 0.003$ the system still reaches the full localization, $\mathcal{L}\rightarrow 0$ for disorder $V\rightarrow -\infty$, at the same concentration 
$C_C=100(n/2)\%$ (for $V<0$). For stronger harmonic confinement, $k>0.003$, we see however that entanglement does not have a minimum at $C_C$: it saturates instead at a finite value $\mathcal{L}\sim0.25$ for $C$ slightly larger than $C_C$. A shift on the critical concentration for full localization could be related to the effective density in harmonic traps \cite{vvfFFLO}, defined as the density at the center of the harmonic potential, which increases with $k$. But the fact that $\mathcal{L}$ saturates at finite values, suggests that for $k>0.003$ the system is actually ordinarily localized. This is confirmed by the ground-state energy shown in Fig.\ref{fig5}-b: there is a signature of the fully-localized state at $C_C$ for all the curves, except for $k=0.004$.

In Figure \ref{fig5}-c we show the entanglement as a function of the harmonic potential strength $k$ for $C<C_C$, $C=C_C$ and $C>C_C$. We observe that for $C\neq C_C$ entanglement saturates with $k$, but never reaches zero, indicating that the system is ordinarily localized. For $C=C_C$, $\mathcal{L}\sim 0$ for weak harmonic traps, but there is a certain maximum confinement (in this case $k^{max}\sim 0.003$) for which the system still fully localizes. For $k>k^{max}$ the system is driven to an ordinary localization, with $\mathcal{L}\neq 0$.

\section{Conclusion}

In summary we have investigated the impact of density, interaction and harmonic confinement on the SIT. We find that the critical 
concentration at which the fully-localized state occurs does not depend on the interaction strength, but requires a minimum disorder 
intensity to appear and depends on the average density with simple relation for both, attractive and repulsive disorder, as given by 
Eq.(\ref{eq1}) and Eq.(\ref{eq2}). When dealing with harmonically confined disordered systems, the full localization is reached at the same critical concentration for weak confinements. But as the confinement increases, the system undergoes a transition from the fully-localized state to the ordinary localization.

\begin{center}\bf ACKNOWLEDGMENTS \end{center}

VVF was supported by FAPESP (13/15982-3). GAC was supported by the Coordena\c{c}\~{a}o de Aperfei\c{c}oamento de Pessoal de Nivel Superior - Brasil (CAPES) -
Finance Code 001.


\end{document}